\documentstyle[aps,prb,twocolumn,psfig]{revtex}
\draft
\def\beq{\begin{equation}}
\def\eeq{\end{equation}}
\def\beqarr{\begin{eqnarray}}
\def\eeqarr{\end{eqnarray}}

\input epsf.tex

\begin{document}
\draft

\twocolumn[\hsize\textwidth\columnwidth\hsize\csname @twocolumnfalse\endcsname

\title{Disorder from disorder in a strongly frustrated transverse field Ising chain}
\author{D. J. Priour$^{1}$, M. P. Gelfand$^{2}$, and S. L. Sondhi$^{1}$}
\address{$^1$Department of Physics, Princeton University, Princeton, New Jersey 08544}
\address{$^2$Department of Physics, Colorado State University, Fort Collins, Colorado
80523}

\date{\today}

\maketitle
\begin{abstract}
We study a one-dimensional chain of corner-sharing triangles with antiferromagnetic
Ising interactions along its bonds. Classically, this system is highly frustrated
with an extensive entropy at $T=0$ and exponentially decaying spin correlations. We 
show that the introduction of a quantum dynamics via a transverse magnetic field
removes the entropy and opens a gap, but leaves the ground state disordered at all
values of the transverse field, thereby providing an analog of the ``disorder
by disorder'' scenario first proposed by Anderson and Fazekas in their search for
resonating valence bond states. Our conclusion relies on exact diagonalization
calculations as well as on the analysis of a 14th order series expansion about the
large transverse field limit. This test suggests that the series method could be used
to search for other instances of quantum disordered states in frustrated transverse
field magnets in higher dimensions.
\end{abstract}

]
%
%
\section{Introduction}

The study of quantum versions of classically frustrated magnets has been a
subject of interest at least since the work of Anderson and Fazekas on the
possibility of a quantum disordered state for the triangular lattice Heisenberg
antiferromagnet.\cite{pwa-faz} Following their work, and especially after the 
early suggestion
of Anderson that the cuprate superconductors derive their special properties
from the proximity of a spin liquid state,\cite{pwa-sc87} there has been a 
considerable amount of work
on quantum Heisenberg models on various frustrated lattices such as the 
kagom\'e.\cite{kagrefs}

One can think of quantum Heisenberg models as (classical) Ising models perturbed
by a transverse (XY) exchange---indeed, this was the strategy followed by Anderson
and Fazekas in their analysis of the triangular lattice problem. Phrased in this
fashion, the problem becomes one of the effects of introducing a quantum dynamics
into a highly degenerate ground state manifold of the Ising system---a procedure
of evident interest on account of the singular effects of {\it any} perturbation.
The 
simplest instance of this more general problem is the introduction of a transverse
magnetic field, which has been used in several contexts previously to generate the 
simplest quantum statistical mechanics; for a review see 
Ref.~\onlinecite{Chakrabarti96} which also
reviews some work on some one-dimensional systems with competing interactions.

A number of geometrically frustrated quantum Ising systems have been studied recently 
by Moessner, Chandra and one of the present authors who have reported instances of 
``order by disorder''\cite{villain} in
which a non-trivial ordering pattern is selected by the quantum fluctuations as
well as one of ``disorder by disorder'' (on the kagom\'e lattice) in which the 
Anderson-Fazekas scenario of a disordered quantum state constructed out of a 
disordered classical manifold is realized.\cite{mcs}  
In this paper, we report studies of a one-dimensional, geometrically
frustrated chain which also exhibits disorder by disorder. This is interesting
in itself; additionally, through a technical advance, we were able to provide
strong evidence for the quantum disordered state by means of a strong coupling
perturbation expansion, which we anticipate will be a useful technique
in studying higher-dimensional frustrated quantum Ising systems. 

\section{The model and its possible ordering}
\label{sec:model}

\begin{figure}
\begin{center}
\centerline{\psfig{figure=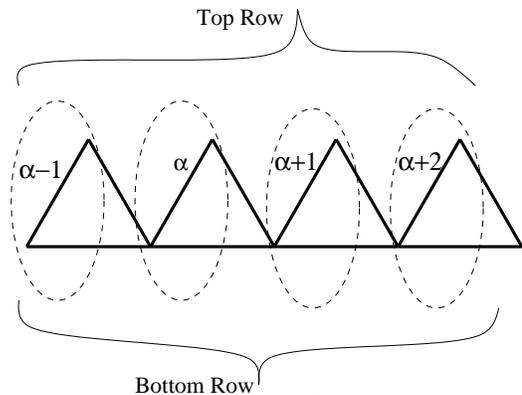,width=2.7in}}
\caption{The triangular Ising spin chain. The dashed ellipses enclose unit cells.}
\label{fig:saw}
\end{center}
\end{figure}

We consider the Hamiltonian
\beq
H = J \sum_{\langle ij \rangle} \sigma_i^z \sigma_j^z + \Gamma \sum_i \sigma_i^x 
\eeq
where $J>0$ is the antiferromagnetic Ising exchange, the sum in the first term
runs over the bonds of the triangular chain in Figure~\ref{fig:saw}, 
$\Gamma$ is the strength of the transverse field and the $\sigma^a$ are the Pauli 
operators for $S=1/2$. (As the true spin operators are $S^a = \sigma^a/2$, when 
$\hbar=1$, our $J$ is really one quarter the exchange and our $\Gamma$ is half
the physical transverse field.)
 
\begin{figure}
\begin{center}
\centerline{\psfig{figure=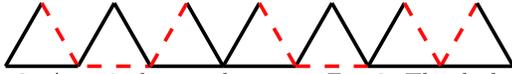,width=2.7in}}
\caption{A typical ground state at $\Gamma=0$.  The dashed lines represent 
frustrated or ``bad" bonds, of which there is exactly one per triangle.}
\label{fig:grnd}
\end{center}
\end{figure}

The ground states of the system without the transverse field can be obtained by
minimizing the energy of each triangle separately and this in turn requires that
we choose the one bond in each that will remain unsatisfied (the ``bad'' bond).
The choice of the bad bonds is totally independent from triangle to triangle (the
system is a bad bond paramagnet); a typical state is exhibited in Fig~\ref{fig:grnd}.
With free 
boundary conditions, the number of such states is thus $2 \times 3^N$ where
$N$ is the number of triangles and the extra factor of $2$ accounts for the two
Ising reversed states that give rise to the same bond configuration.\cite{fn-periodic}
This extensive entropy is accompanied by short-ranged correlations and we find that
the spin-spin correlation function averaged over the ground state manifold takes
the form
\beqarr
\langle S_{\alpha t}^z S_{\beta t}^z \rangle = \langle S_{\alpha b}^z S_{\beta
t}^z \rangle &=& (-1/3)^{\beta-\alpha+1}\\
\langle S_{\alpha b}^z S_{\beta b}^z \rangle = \langle S_{\alpha t}^z S_{\beta
b}^z \rangle &=& (-1/3)^{\beta-\alpha}
\eeqarr
where $\alpha t$ and $\alpha b$ denote the top and bottom spins in unit cell $\alpha$.
Due to the lack of inversion symmetry in our choice of unit cell, these forms hold
for $\beta > \alpha$; they can be extended to cover $\beta \le \alpha$ by inspecting
the lattice and noting its symmetries.
Evidently, the system is disordered at all temperatures.

We turn now to the quantum problem posed by the inclusion of the transverse field. 

\subsection{Analysis at $\Gamma \ll J$}

\begin{figure}
\begin{center}
\centerline{\psfig{figure=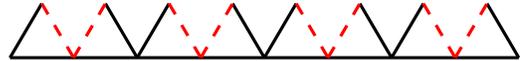,width=2.7in}}
\caption{The maximally flippable configuration. All sites along the top
row and half of those along the bottom row are flippable.}
\label{fig:nixon}
\end{center}
\end{figure}

At $T=0$, an infinitesimally small $\Gamma$ is a singular perturbation which will
lift the macroscopic degeneracy of the classical problem.
Following Ref.~\onlinecite{mcs}, we will attempt
to identify the physics of this regime variationally. To this end we identify the
configuration (unique, up to global symmetry operations) that maximizes the number of
``flippable'' spins, which is shown in Figure~\ref{fig:nixon}. A flippable
spin in a given configuration is one which can be reversed without violating
the ground state constraint. As a flippable spin can be polarized along
$\hat{x}$ at no cost in exchange, we expect a state that maximizes their number to
be especially favored by the transverse field. As not all flippable spins are independently
flippable, we have two options. We may construct the {\em staggered} state in which we
polarize the maximal set of independently flippable spins along the top row,
\beq
| s \rangle = (\otimes_{\rm top} |x-\rangle) \  | \uparrow \downarrow \uparrow 
\downarrow \uparrow \downarrow \cdots \rangle_{\rm bottom} \ ;
\eeq
here $|x-\rangle$ is the state with a spin pointing down along the x axis. 
Alternatively we may construct the {\em uniform} state in which we 
polarize all the flippable spins but
correct for the conflicts by projecting onto the ground state manifold
by the action of the projector $P$ which eliminates those configurations
that do not have exactly one bad bond per triangle:
\beqarr
| u \rangle &=&  P \left\{ (\otimes_{\rm top} |x-\rangle) \right. \\
& & \qquad \otimes \left.  (| \uparrow\rangle \otimes
|x-\rangle \otimes  |\uparrow\rangle \otimes |x-\rangle 
\otimes  |\uparrow\rangle \cdots )_{\rm bottom} \right\}  . \nonumber 
\eeqarr
The uniform state inherits all the symmetries of the maximally flippable configuration
but the staggered state does not. For the wave functions as written, we can readily 
evaluate the energies and we find that the staggered state has the same energy 
$-(\Gamma + J)/2$ per site as the uniform state. We can also write down the magnetization
(up to translations and Ising reversal) in the staggered state:
\beq
\langle \sigma^z(\alpha) \rangle = m(\alpha) = \left\{ \begin{array}{ll}
                     0   & {\rm top} \\
                   (-{1 \over 2})^\alpha  & {\rm bottom}
                   \end{array}
                   \right.
\eeq
and the uniform state
\beq
m(\alpha) = \left\{ \begin{array}{ll}
                     -{1 \over 10}   & {\rm top} \\
                   {1 \over 10} + (-{2 \over 5})^\alpha  & {\rm bottom \ .}
                   \end{array}
                   \right. 
\eeq
Needless to say, this analysis being purely variational is only indicative of what
kind of ``order from disorder'' the system {\em might} exhibit at $\Gamma \ll J$.
At issue is whether the connectivity of the set of configurations is too high
(e.g. as reflected in the disorder in the classical problem) to permit a 
localization of the ground state wavefunction near the maximally flippable 
configuration. 
As advertised we will, instead, find that the system remains disordered.

\subsection{Analysis at $\Gamma \gg J$} 

One of the attractive features of transverse field Ising models, emphasized
in Ref.~\onlinecite{mcs}, is that they exhibit a gapped, paramagnetic
phase at $\Gamma \gg J$. This enables a natural perturbative expansion
for the problem, in which we perturb in the Ising exchange about the purely
transverse field paramagnet. A second attractive feature, exhibited by
the well known Trotter-Suzuki procedure\cite{suzuki}  is that the quantum
partition function (with the transverse field present) has the form of the 
classical partition function of copies of the chain (now without the
transverse field) coupled {\em ferromagnetically} in the imaginary time
direction which can then be analyzed from its ``high temperature'' phase
via the standard Landau-Ginzburg-Wilson (LGW) procedure.\cite{blank}  It is not
difficult to show that these are attempts to search for the same instability.
We will pursue the perturbative expansion a bit later; here, we will
attempt to identify the leading instability from the paramagnetic phase
by the LGW step of diagonalizing the bond (adjacency) matrix. (The
equivalence to the first step in the perturbation expansion will be evident when
we come to it.)

As the ferromagnetic interaction in the imaginary time direction is trivially
incorporated by including a quadratic dispersion about $k_\tau=0$, we
focus on the problem in the chains. To diagonalize the quadratic form,
\beq
{\mathcal H} = \sum_{\langle ij \rangle} S_i S_j
\eeq 
we introduce Fourier variables with respect to the unit cell location, 
\begin{eqnarray}
\label{eq:1.1eq}
S_{\alpha t} &=& \sum_{k} e^{ik \alpha} S_{k t}\\
\label{eq:1.2eq}
S_{\alpha b} &=& \sum_{k} e^{ik \alpha} S_{k b} \ .
\end{eqnarray}
and find that the Hamiltonian takes the form
\begin{eqnarray}
{\mathcal H} &=& 
\label{eq:1.7eq}
\sum_{k>0} \left[ \begin{array}{lr} S^{*}_{k t} & 
S^{*}_{k b} \end{array} \right] \left[ \begin{array}{lr} 0 & (1 + e^{ik}) \\
(1 + e^{-ik}) & 2\cos k \end{array} \right] 
\left[ \begin{array}{c} S_{k t}\\ S_{k b} \end{array} \right] \ .
\end{eqnarray}
Further diagonalization yields the eigenvalues 
\begin{equation}
\label {eq:1.8eq}
\lambda_{\pm} = (\varepsilon - 1) \pm \sqrt{\varepsilon^{2} + 1}
\end{equation}
where we have defined $\varepsilon  = 1+ \cos k$. 
Evidently, $\lambda_{-}$ is minimized when one has $\varepsilon=0 \rightarrow k = \pi$.  
The corresponding eigenvector is 
\begin{equation}
\label{eq:1.9eq}
\vec{v}_{\rm min} = \left[ \begin{array}{c} 0 \\ 1 
\end{array} \right]
\end{equation}
Hence, the spatial dependence of the lowest eigenmode, which is the candidate for
ordering, is 
\begin{equation}
\label{eq:1.10eq}
\phi_{\pi} = \cos \pi\alpha \left[ \begin{array}{c} 0 \\ 1 \end{array} \right] =
 (-1)^{\alpha} \left[\begin{array}{c} 0 \\ 1 \end{array} \right] \ .
\end{equation}
Thus, as $\Gamma/J$ is lowered---which corresponds to lowering the temperature
in the classical representation---the LGW analysis suggests a transition in the 
$d=2$ Ising universality class\cite{fn-isingtrans} to a state with the top row 
disordered and bottom row ordered antiferromagnetically, which is the 
staggered state that we had constructed earlier.
 
\subsection{Mean Field Theory}

As we are interested in the suppression of ordering by fluctuations
due to the frustration, it is useful to have a mean-field estimate
for the location of the transition. To this end we consider the
possibility of ordering into the staggered state. In this case only
the spins on the bottom row see an effective field and the problem
(in mean field theory) is identical to that of the purely one-dimensional
transverse field Ising chain. This yields a critical coupling 
\begin{equation}
\label{eq:1.16eq}
x_c =  (J/\Gamma)_c = 1/2 
\end{equation}
and a staggered magnetization
\beq
m = \sqrt{1 - (2x)^{-2}}
\eeq
in the ordered phase. The magnetization of the top row is always zero.
Better estimates can be obtained by treating the top row in mean field
theory and using the known exact results for the bottom row.\cite{kogut}  
In this fashion we obtain $x_c=1$ and $m \sim (x-x_c)^{1/8}$.

\section{Exact diagonalization}

\subsection{Modified Lanczos method}
We now turn to our numerical studies of the model, beginning with
exact diagonalization of the Hamiltonian for finite lattices with
periodic boundary conditions.
Exact diagonalization unfortunately imposes
a demand on computer memory which rises exponentially with the size of the
system considered. Consequently, the systems discussed in this 
report are limited to at most 8 unit cells (16 spins).

In the modified Lanczos method that we apply,\cite{dagotto}  
one begins with a state $| 0 \rangle$ 
expected to have nonzero overlap with the ground state of $\mathcal{H}$.
For the present problem we used the trivial (fully $x$-polarized)
$J=0$ ground state, which is a $q=0$ state.
One generates a state orthogonal to $| 0 \rangle$ by acting upon $| 0 \rangle $ with
$\mathcal{H}$ and using Gram-Schmidt orthogonalization to generate      
the state
\begin{equation}
\label{eq:1.17eq}
|1 \rangle = {\mathcal H} | 0 \rangle - \frac{\langle 0 | {\mathcal H} | 0 \rangle}
{\langle 0 | 0 \rangle}
\end{equation}
One then diagonalizes $\mathcal{H}$ in the resulting two-dimensional
subspace.  The lowest eigenvalue is the improved estimate for the 
ground state energy, and the corresponding eigenvector is the improved 
estimate for the ground state wave function.
This process is iterated to convergence.

A slight modification of the procedure used in obtaining the
ground state energy and wave function can be employed to find the
energy gap and first excited state wave function.
One begins with an initial guess with nonzero
overlap with the lowest excited state wave function and orthogonal
to the ground state, and then proceeds as described above
for the ground state.
The initial guess for the excited state was chosen to 
be one of the excited states in the $J=0$ system,  with a 
single flipped spin at a particular location.  Although the state is not 
one of definite momentum, many iterations of the modified Lanczos 
procedure converge to a state of definite momentum in which 
$k = \pi$, as expected from the analysis in Section~\ref{sec:model}.   
Although successive excited state estimates should in principle  
be orthogonal to the ground state, rounding errors cause an     
admixture with the ground state to occur at each iteration.  
To prevent a convergence toward the ground state, Gram-Schmidt 
orthogonalization is used at each step to remove any component of 
the ground state which may have entered.  Although one does not 
have the exact ground state at a particular iteration, convergence 
can still be achieved by performing Gram-Schmidt orthogonalization 
with the most recent estimate for the ground state wave function. 

\subsection{Excitation gap}

\begin{figure}
 \begin{center}
\centerline{\psfig{figure=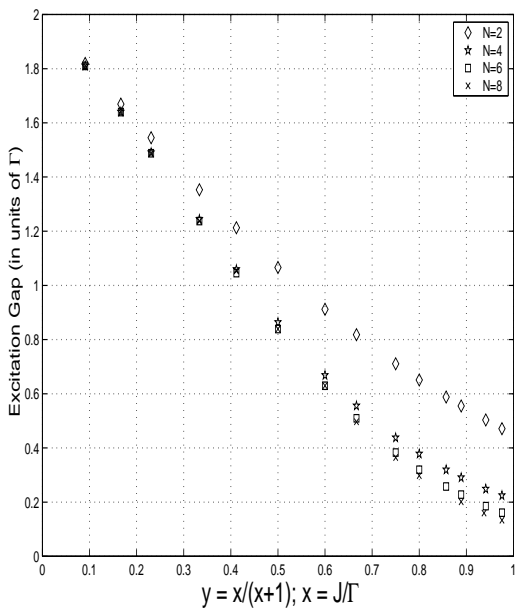,height=3.2in,width=2.7in}}
\caption{Gaps obtained via exact diagonalization for $N = 
{2,4,6,8}$}
 \label{fig:langap}
 \end{center}
 \end{figure}

Figure~\ref{fig:langap} displays the gap to the first excited state
computed for a variety of system sizes ($N$ denoting the number
of unit cells) and values of $x=J/\Gamma$.  
For convenience, the domain $[0,\infty]$ of $x$
is compressed into the interval $[0,1]$
via the transformation $y = x/(x+1)$.  
For most values of $y$, convergence of the gap with increasing system size
is quite rapid, in itself an indication of the presence of a short correlation
length. In fact, for $y<2/3$, the gaps for systems with $6$ and $8$ unit
cells are virtually indistinguishable.  
In that regime, it is clear that the system remains disordered (in the same
phase as the pure transverse-field model $y=0$).  For larger $y$ the situation
is less clear-cut, so we next consider a more sensitive test for the existence
of a continuous phase transition. (We note that our largest system size yields
a ground state energy of $-\frac{1}{2} \left[ J+ (1.11)\Gamma \right] $ per spin 
at $\Gamma \ll J$, 11\% lower than the energy for the variational states
considered earlier.)

\subsection{Cumulants}

Following Binder,\cite{binder} we examine the cumulants

\begin{equation}
\label{eq:1.18eq}
r = \frac{\langle | \Psi |^{4} \rangle}{\langle | \Psi |^{2}  
\rangle ^{2}} 
\end{equation}
In equation~\ref{eq:1.18eq}, $\Psi$ is the order parameter of interest.
Since any ordering is expected to be accompanied by  
antiferromagnetic ordering of the bottom row, a sensible choice of 
$\Psi$ seems to be the staggered magnetic moment operator, 
\begin{equation}
\label{eq:1.19eq}
M_s = \sum_{\alpha} (-1)^{\alpha} \sigma_{\alpha b}^{z}
\end{equation}
on the bottom row.

It is expected that for systems of size $L$, in the vicinity
of a critical point, $r$ obeys the scaling form 
\begin{equation}
r = f((x- x_{c})L^{1/\nu})
\end{equation}
If a critical point exists, the cumulants computed for various system
sizes should intersect at the same point for $x = x_c$.  

\begin{figure}
 \begin{center}
\centerline{\psfig{figure=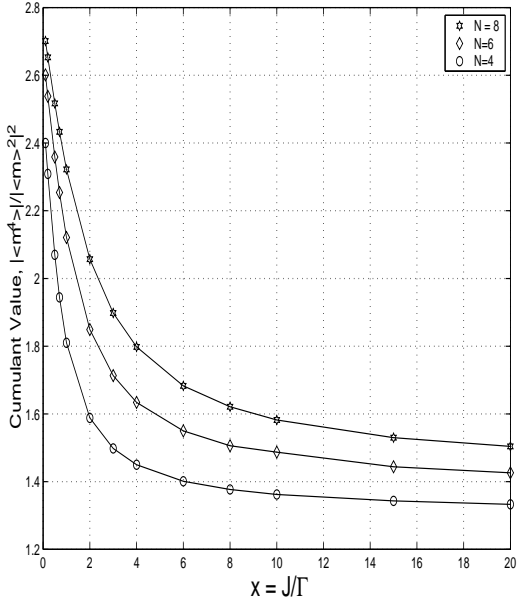,height=3.2in,width=2.7in}}
\caption{Plots of the cumulant $\langle M_s^{4} \rangle/
| \langle M_s^{2} \rangle |^{2}$ for triangular chains containing 
$4$, $6$, and $8$ unit cells.}
 \label{fig:cumula}
 \end{center}
 \end{figure}

Cumulants for systems containing $4$, $6$, and $8$ unit cells are 
displayed in Figure~\ref{fig:cumula}.  There is apparently no intersection of the 
cumulants for different system sizes for $x\leq20$ (and, by extrapolation, for
even larger values of $x$).  The behavior of the cumulants is consistent
with the absence of a critical point and therefore suggests that the 
system is disordered for all values of $J/\Gamma$. We note that in a
disordered phase with a Gaussian distributed order parameter, the
cumulant will approach the value 3. At $x=0$ the cumulant can be shown
analytically to have the value $3 - 4/N$ for $N$ unit cells; evidently
the finite size effects are greater at large $x$, where the ground state
constraint produces correlations, but even there the steady growth with
system size is consistent with a thermodynamic limit value of 3. 

\section{Perturbation expansion generation and analysis}

\subsection{Cluster expansions for the elementary excitation spectra}

A perturbation expansion for the energy gap (and the
full elementary excitation spectrum, incidentally)
about the $J=0$ limit was constructed to 14th order
in $J/\Gamma$ using a cluster expansion technique.
The ideas underlying this calculation are set out
briefly in Ref.~\onlinecite{gelfand96}, and described
in more detail in Ref.~\onlinecite{gelfand00}.
However, this particular calculation is unusual 
(but not unique, see M\"uller-Hartmann {\it et al.}\cite{MH99})
in one interesting respect: it is carried out for a system with 
more than one branch of elementary excitations which are degenerate
in the unperturbed limit, but are not degenerate by symmetry
and hence are not degenerate for $J>0$.
That such calculations can be done by cluster expansion methods
was mentioned briefly in Ref.~\onlinecite{gelfand00}.  
Here we supply some further details
which are important in making the connection between the
the immediate product of the cluster expansion, which is a
matrix containing power expansions that describe the 
motion of elementary excitations in real space, and the 
desired end product, the power series expansions for the
excitation spectra in wave vector space.

For the purposes of this section we assign the sites at the
base of the triangles the coordinates $(n,0)$, and those at the
tips of the triangles $(n+{1\over2},1)$, with $n$ running over the integers.
The unperturbed Hamiltonian is the transverse field
$\sum_n [ \sigma_{(n,0)}^x + \sigma_{(n+{1\over2},1)}^x ]$
and the perturbation is the Ising exchange
$\sum_n [ \sigma_{(n,0)}^z\sigma_{(n+1,0)}^z 
+\sigma_{(n,0)}^z\sigma_{(n+{1\over2},1)}^z
+\sigma_{(n+1,0)}^z\sigma_{(n+{1\over2},1)}^z ]$.
The perturbation expansion is a power series in $x=J/\Gamma$.

For the unperturbed system, the ground state consists of all spins ``down"
(pointing along $-x$)
and the elementary excitations are single spins which
have been flipped ``up''.  Note that there are two degenerate branches
of such excitations, flat bands at $\omega=2$. (One could consider an 
unperturbed Hamiltonian of the more general form 
$\sum_n \left[ \sigma_{(n,0)}^z + \zeta \sigma_{(n,1)}^z \right]$, which has
the same symmetry as the triangular chain but lacks the degeneracy.
For $\zeta\neq1$, then, perturbations expansions for each of
the branches of excitations could be constructed separately;
however, there would be energy denominators $(1-\zeta)$ in abundance
that would lead to poorly converging series for $\zeta$ near 1.)

In our calculations, with $\zeta=1$, we treat all of the 
elementary excitations on equal footing.    We keep track of
three distinct types of matrix elements in the effective
Hamiltonian, those that connect two base sites (type $A$),
those that connect two tip sites (type $B$), and those
that connect base to tip sites (type $C$); let us call
$t^{X}_d$ the matrix element (which is a series expansion
in $\lambda$) in the effective Hamiltonian of type $X$ that 
couples sites at horizontal distance $d$.  The effective
Hamiltonian has the following structure near the diagonal
\begin{equation}
\left(
\begin{array}{cccccccc}
\ddots & \vdots & \vdots & \vdots & \vdots & \vdots & \vdots & \\
\cdots & t^{A}_0 & t_{1/2}^{C} & t_1^{A} & t_{3/2}^{C} & t_2^{A} & t_{5/2}^{C} &
\cdots \\
\cdots & t_{1/2}^{C} & t^{B}_0 & t_{1/2}^{C} & t_1^{B} & t_{3/2}^{C} & t_2^{B} &
\cdots\\
\cdots & t_1^{A} & t_{1/2}^{C} & t^{A}_0 & t_{1/2}^{C} & t_1^{A} & t_{3/2}^{C} & 
\cdots\\
\cdots & t_{3/2}^{C} & t_1^{B} & t_{1/2}^{C} & t_0^{B} & t_{1/2}^{C} & t_1^{B} & 
\cdots\\
\cdots & t_2^{A} & t_{3/2}^{C} & t_1^{A} & t_{1/2}^{C} & t^{A}_0 & t_{1/2}^{C} & 
\cdots\\
\cdots & t_{5/2}^{C} & t_2^{B} & t_{3/2}^{C} & t_1^{B} & t_{1/2}^{C} & t_0^{B} & 
\cdots\\
 & \vdots & \vdots & \vdots & \vdots & \vdots & \vdots & \ddots 
\end{array} 
\right).
\end{equation}
This matrix acts on a column-vector
describing the spin-flip amplitude at each site, listed in order
of its $x$ coordinate.
This matrix is readily diagonalized by plane waves with a two-site
basis. One obtains the two branches of the excitation spectrum in the form
\begin{equation}
\label{eq:quadratic}
\epsilon_{\pm}={1\over2}\left(
F_A + F_B \pm \sqrt{(F_A - F_B)^2 + F_C^2}
\right)
\end{equation}
with 
\begin{equation}
F_A(q)=\sum_{n\ge0} \tilde t^{A}_n \cos nq \  ,
\end{equation}
likewise for $F_B(q)$, and
\begin{equation}
F_C(q)=\sum_{n\ge0} \tilde t^{C}_{(2n+1)/2} \cos (2n+1)q/2 \ .
\end{equation}
The relationship between the $t$s and $\tilde t$s is as follows.
Take any two adjacent rows of the effective Hamiltonian, and
count how many $t^{X}_d$ there are for a given $X$ and $d$:
that number is the ratio $\tilde t^{X}_d/t^{X}_d$.
Thus $\tilde t^{A}_0=t^{A}_0$, $\tilde t^{C}_{1/2}=4t^{C}_{1/2}$,
and so forth.  It is the
$\tilde t$s that come directly out of our computer programs and so
it is convenient to express the functions that appear in
Eq.~(\ref{eq:quadratic}) in those terms.

We should note that our 14th order calculation of the excitation
spectrum involved weight calculations for 1355 graphs of up to
15 spins.  We did not attempt to classify the graphs topologically.
The implementation of the weight calculations traded off
considerable efficiency for safety (guarding against a variety
of possible coding errors), and the calculations were carried
out on a modest machine (333 MHz Pentium II with 256 megabytes of RAM),
so it should be feasible to evaluate several more terms of the
perturbation expansion if desired.

\subsection{Analysis of the energy gap series}

One is faced immediately with an interesting choice in
analyzing the series for the gap, $\epsilon_-$.
One can either extrapolate the series for $F_A$, $F_B$ and
$F_C$ individually and insert the results into the formula
(\ref{eq:quadratic}), or evaluate that formula for the series
and extrapolate $\epsilon_-$ directly.            

However, the lowest excited state is always found at $q = \pi$, 
and in this case the need to choose vanishes because $F_C(\pi)\equiv0$
and thus $\epsilon_-(\pi)=\min(F_A(\pi),F_B(\pi))=F_A(\pi)$.
In Table~\ref{tab:1tab} we display $F_A(\pi)$ and $F_B(\pi)$.
The complete set of $\tilde t_D^X$ are available from the authors
on request.

\begin{table}
\caption{Coefficients of $x^n$ in the series $F_A(\pi)$ and $F_B(\pi)$.}
\label{tab:1tab}
\begin{tabular}{cdd}
 $n$ & $F_{A}(\pi)$ & $F_{B}(\pi)$ \\
\tableline
 0  &    2.000000000 &    2.000000000 \\
 1  & $-2$.000000000 &    2.000000000 \\
 2  &    1.000000000 &    1.000000000 \\
 3  &    0.000000000 & $-0$.750000000 \\
 4  & $-0$.250000000 &    0.000000000 \\
 5  &    0.093750000 & $-0$.328125000 \\
 6  & $-0$.144531250 &    0.785156250 \\
 7  &    0.396484375 & $-0$.621093750 \\
 8  & $-0$.544921875 &    0.763671875 \\
 9  &    0.632377624 & $-1$.323387146 \\
 10 & $-0$.908432960 &    1.746130307 \\
 11 &    1.405481246 & $-2$.362823632 \\
 12 & $-2$.073822043 &    3.583402837 \\
 13 &    3.127851625 & $-5$.491168599 \\
 14 & $-4$.983237137 &    8.550544610
\end{tabular}
\end{table}

\begin{figure}
\begin{center}
\centerline{\psfig{figure=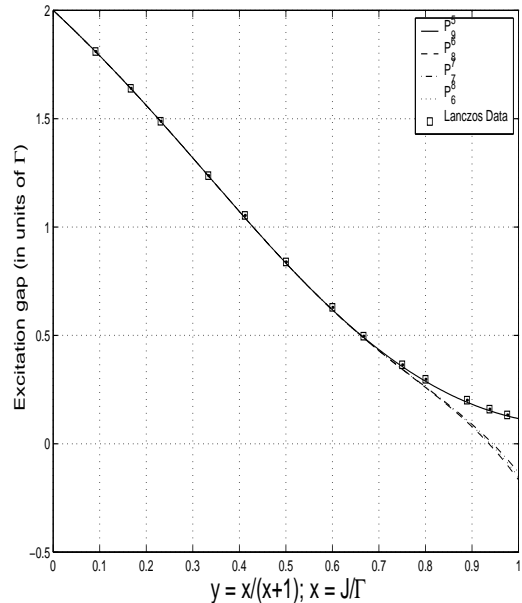,height=3.2in,width=2.7in}}
\caption{Pad\'{e} approximants and exact diagonalization results ($N=8$) for the 
gap plotted as a function of $y=x/(x+1)$.}
\label{fig:pade}
\end{center}
\end{figure}

To analyze the gap series, we applied
the transformation $y=x/(x+1)$, and constructed Pad\'e
approximants to the transformed series.
We use the notation $P^M_{M'}$ to denote an approximant with a polynomial
of order $M$ in the numerator and order $M'$ in the denominator.
Figure~\ref{fig:pade} displays several approximants which use
all the terms of the series (and hence $M+M'=14$).

A salient feature of Figure~\ref{fig:pade} is the consistency of the Pad\'{e} 
approximants over a large fraction of the domain. 
The approximants displayed in the plot are consistent to within 
$1\%$ for $y$ as large as $0.75$. Note that this corresponds to a value of
$x=3$, well in excess of the mean field estimate of $x_c=1/2$ and the $d=1$
transverse field Ising model estimate of $x_c=1$. Even without further
analysis these indicate the disordering  effects of the strong frustration
in the problem. 

Figure~\ref{fig:pade} also displays the excitation gap 
as computed via the modified Lanczos algorithm.
Evident in the graph is the excellent 
agreement of the approximants and the exact diagonalization results 
over the region of consistency among the Pad\'{e} approximants.

\subsubsection{Two Point Approximants and The Global Phase Portrait}

Although the region of validity of the analyzed gap series extends  
well into the domain of strong exchange coupling, the Pad\'{e} approximants
plotted in Figure~~\ref{fig:pade} tell an ambiguous story.  In 
particular, it is not clear from the displayed approximants whether or not the 
excitation gap vanishes for large $J/\Gamma$ and hence it is unclear 
whether the triangular chain is ordered in the $J\gg\Gamma$ limit.  
The reader should note though that even those Pad\'{e} approximants
that indicate a vanishing of the gap, do so for values of $x$ in excess
of 11.

\begin{figure}
\begin{center}
\centerline{\psfig{figure=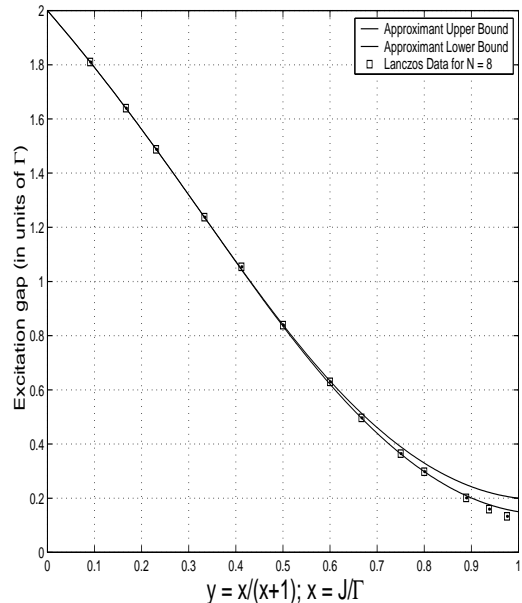,height=3.2in,width=2.7in}}
\caption{Biased Pad\`{e} approximant $P^{5}_{9}$ plotted at 
both limits of $1\%$ consistency in juxtaposition with Lanczos results
for $N=8$.}
\label{fig:twopoint}
\end{center}
\end{figure}

Biasing of the Pad\'{e} approximants is one means of extending the results
of the gap series analysis to larger $y$ values.  A simple way to bias the 
approximants is by means of
``two-point'' approximants.\cite{bender-orszag}   
In particular, one would hope that by biasing the value of
the gap at $y=1$ the resulting approximants would be accurate
over the entire range $0\leq y \leq 1$.
However, since it is precisely the value of the gap at $y=1$ (let us
denote that $\Delta(1)$) that
we know least well, such a procedure appears to beg the question.
What we have done, then, is to construct two-point approximants (with
various orders of numerators and denominator) for
a range of $\Delta(1)$ values, and observed behavior.
Two of 
the biased approximants, $P^{5}_{9}$ and $P^{6}_{8}$, show a high degree of consistency with each other for a relatively small 
range of $\Delta(1)$ values.   
  In fact, the range over which the maximum
discrepancy between the two is less than $1\%$ is confined between 
$\Delta(1) = .15$ and $\Delta(1) = .2$.
For values of $\Delta(1)$ between $.15$ and $.20$, $P^{5}_{9}$ and 
$P^{6}_{8}$ are in good enough agreement that the maximum disagreement between
the two over the entire $y$ domain falls below $1\%$.  Outside the $\Delta(1)$
range indicated above, the maximum disagreement between $P^{5}_{9}$ and $P^{6}_{8}$
exceeds $1\%$.  Figure~\ref{fig:twopoint}
displays the biased approximants corresponding to the limits of $\Delta(1)$      
discussed above.  
    No matter what the choice of $\Delta(1)$, none of the other two point approximants are in agreement with each other or with $P^{5}_{9}$ and $P^{6}_{8}$ 
to an extent which approaches the level of agreement of 
 $P^{5}_{9}$ and $P^{6}_{8}$ for $\Delta(1)$ values
between $.15$ and .$.20$.

The biased  $P_{9}^{5}$ approximants displayed in the figure are in reasonable, though 
not perfect agreement with the Lanzcos gap values.  The approximants shown 
are biased at $\Delta(1) = .15$ and $\Delta(1) = .2$, the bounds of the 
region in which $P^{5}_{9}$ and $P^{6}_{8}$ satisfy the $1\%$ consistency
criterion discussed above.  The agreement is 
best for small values of $y$ but deteriorates as $y$ approaches 1.  However, 
even at $y = 1$, averaging  the two approximants yields 
a prediction for $\Delta(1)$ of $.175$, a value which differs from the 
Lanczos prediction of $.132$ (albeit without any attempt at scaling with
system size) by only $30 \%$. The approximants 
echo the exact diagonalization results in indicating the absence of order 
for all values of $J/\Gamma$.

\section{Summary}

Results for the energy gap and cumulants from exact diagonalization of finite systems 
indicate that the transverse field, triangular chain Ising model is 
disordered, even though several analytic approaches (from both the small
and large transverse-field limits) suggest that a particular ordered phase could
exist for small transverse field. We note that the lowest lying excited state
is consistent with the ordering analysis, indicating that the fluctuations
generated by the ground state constraint at small $\Gamma/J$ are too strong to allow 
order to set in. Hence we get ``disorder by disorder'' instead of ``order by
disorder''.
Estimates of the gap obtained by direct Pad\'e approximants to 14th order
perturbation expansions about the strong transverse field limit
produce excellent results over a modest range of the Ising 
exchange coupling to transverse field ratio that exceeds the natural
estimates for a critical value by a factor of 3 and do not indicate ordering
for values of the ratio as big as 11.  Globally 
reasonable results have been obtained by biasing the approximants 
and applying a consistency criterion and these indicate a lack of ordering
at {\it any} value of $J/\Gamma$.  
This suggests that an application of this method
to systems of greater experimental interest, 
such as the higher dimensional kagom\'e and pyrochlore lattices, may be a 
viable approach to deducing the global phase behavior of such systems.  
We should note that in $d>1$ we expect the series method to be increasingly
competitive with exact diagonalization as the latter technique has to
contend with much stronger finite size effects at computationally feasible 
system sizes.

\acknowledgments
We would like to thank Roderich Moessner for collaboration in the initial
stages of the project and for much useful advice in its course. We would
also like to thank him for a careful reading of the manuscript.
This work was supported by the National Science Foundation through
grants DMR 94-57928 (MPG) and DMR 99-78074 (DJP and SLS) as well as by the
Alfred P. Sloan Foundation and the David and Lucille Packard Foundation
(SLS).


\begin{thebibliography}{99}
\bibitem{fn-periodic} With periodic boundary conditions there are $3^N + (-1)^N$
states where $N$ is the number of unit cells. Hence we have the same entropy
per site in the thermodynamic limit. The reduction by a factor of two arises
for the periodicity requires an odd/even number of bad bonds along the bottom
row for an odd/even number of unit cells.
\bibitem{fn-isingtrans} The transverse field ferromagnetic and antiferromagnetic
Ising problems are identical and in the same universality class as the
d=2 classical Ising model. As the LGW analysis indicates an antiferromagnetic
ordering, the conclusion follows.
\end{thebibliography}
\end{document}